\documentclass[preprint,nature]{revtex4-1} 
\usepackage{braket}
\usepackage{amsmath}
\usepackage{bm}
\usepackage{amssymb}
\usepackage{float}
\usepackage{graphicx}
\usepackage{caption}
\usepackage{wrapfig}
\usepackage{subcaption}
\usepackage{epstopdf}
\usepackage{booktabs}
\usepackage{multirow}
\usepackage[dvipsnames]{xcolor}

\newcommand\minus{%
  \setbox0=\hbox{-}%
  \vcenter{%
    \hrule width\wd0 height \the\fontdimen8\textfont3%
  }%
}

\begin{document}
\title{Characterization of methanol as a magnetic field tracer in star-forming regions}
\author{Boy Lankhaar}
\email{boy.lankhaar@chalmers.se}
\affiliation{Department of Space, Earth and Environment, Chalmers University of Technology, Onsala Space Observatory, 439 92 Onsala, Sweden}
\author{Wouter Vlemmings}
\affiliation{Department of Space, Earth and Environment, Chalmers University of Technology, Onsala Space Observatory, 439 92 Onsala, Sweden}
\author{Gabriele Surcis}
\affiliation{Joint Institute for VLBI ERIC, Postbus 2, 7990 AA Dwingeloo, The Netherlands}
\affiliation{INAF, Osservatorio Astronomico di Cagliari Via della Scienza 5, I-09047 Selargius, Italy}
\author{Huib Jan van Langevelde}
\affiliation{Joint Institute for VLBI ERIC, Postbus 2, 7990 AA Dwingeloo, The Netherlands}
\affiliation{Sterrewacht Leiden, Leiden University, Postbus 9513, 2330 RA Leiden, the Netherlands}
\author{Gerrit C. Groenenboom}
\affiliation{Theoretical Chemistry, Institute for Molecules and Materials, Radboud University, Heyendaalseweg 135, 6525 AJ Nijmegen, The Netherlands}
\author{Ad van der Avoird}
\affiliation{Theoretical Chemistry, Institute for Molecules and Materials, Radboud University, Heyendaalseweg 135, 6525 AJ Nijmegen, The Netherlands}
\date{\today}

\maketitle
\textbf{Magnetic fields play an important role
during star formation\cite{crutcher:12}. Direct magnetic field strength
observations have proven specifically challenging in the extremely
dynamic protostellar phase\cite{vlemmings:11, sarma:09, sarma:11}.
Because of their occurrence in the
densest parts of star forming regions, masers, through polarization
observations, are the main source of magnetic field strength and
morphology measurements around protostars\cite{vlemmings:11}. Of all
maser species, methanol is one of the strongest and most abundant tracers of
gas around high-mass protostellar disks and in outflows. However, as
experimental determination of the magnetic characteristics of methanol
has remained largely unsuccessful\cite{jen:51}, a robust magnetic field
strength analysis of these regions could hitherto not be performed. Here
we report a quantitative theoretical model of the magnetic properties of
methanol, including the complicated hyperfine structure that results
from its internal rotation\cite{lankhaar:16}.
We show that the large range in values of the Land\'{e} g-factors of the
hyperfine components of each maser line lead to conclusions which differ
substantially from the current interpretation based on a single
effective g-factor. These conclusions are more consistent with other
observations\cite{baudry:98,wright:04} and confirm the presence of
dynamically important magnetic fields around protostars. Additionally,
our calculations
show that (non-linear) Zeeman effects must be
taken into account to further enhance the accuracy of cosmological
electron-to-proton mass ratio determinations using
methanol\cite{bagdonaite:13b,kanekar:15,jansen:11a,dapra:17}.}

The presence of a magnetic field within an astrophysical maser produces
partially polarized radiation. Linear polarization provides information
on the magnetic field direction, while the magnetic field strength can
be determined by comparing the field-induced frequency shifts between
left- and right-circularly polarized maser emission. Extraction of the
relevant information from polarized maser spectra requires knowledge of
the Zeeman parameters which describe the response of the maser
molecule/atom to a magnetic field. These Zeeman parameters are known for
maser molecules such as OH, H$_2$O, and SiO, but not for methanol.

Various torsion-rotation transitions of methanol have been observed as
astrophysical masers. It has long been known that these transitions have
a hyperfine structure, but only recently has an accurate model of this
structure been presented\cite{lankhaar:16}. It was shown that the
so-called torsional motion of CH$_3$OH about the CO bond drastically
complicates the hyperfine interactions and that `spin-torsion' terms
occur in addition to the usual `spin-rotation' and spin-spin coupling
terms. The magnetic moments that produce the hyperfine structure also
interact with an external magnetic field, a Zeeman effect. Here, we
extend the model of methanol's hyperfine structure with the Zeeman
interactions. We quantitatively determined all the relevant coupling
parameters, including effects from the torsional motion, by
quantum-chemical \textit{ab initio} calculations and an estimate of the
torsional Zeeman effect based on experimental
results\cite{engelbrecht:75}. With this model, we can determine the
Zeeman splitting of the hyperfine states within all the known methanol
maser transitions. 

Zeeman interactions are usually described in a first-order approximation
by the Land\'{e} g-factor. In methanol, each torsion-rotation transition
is actually split into a number of transitions between individual
hyperfine levels of the upper and lower torsion-rotation states (Figure~1). The Land\'{e} g-factors calculated for the different hyperfine
transitions differ strongly and, even though these transitions cannot be
individually resolved in the observed maser spectra, we will show that
this is important for the interpretation of the measured polarization
effects. Furthermore, we found that in several cases the energy gaps
between hyperfine levels are so small that hyperfine states with
different total angular momenta $F$ get mixed even by a weak magnetic
field. In such cases, the first-order approximation for the Zeeman
interactions breaks down, and the Zeeman splittings depend non-linearly
on the magnetic field strength (Figure~2). Also the Einstein
A-coefficients of transitions between two hyperfine levels become
magnetic-field dependent quantities in these cases (Supplementary Figure~2). This behavior has not previously been seen in Zeeman interactions
for non-paramagnetic molecules, and is therefore not accounted for in
current maser-polarization theory\cite{deguchi:90,vlemmings:06}.

To apply our results to maser-polarization measurements, we must
consider hyperfine-specific effects in the maser action. The individual
hyperfine lines are not spectrally resolved, but the maser action
can favor specific hyperfine transitions by the following mechanisms:
i)~varying radiative rates for stimulated emission (see the Einstein
coefficients of the various hyperfine components within a
torsion-rotation line (Supplementary Information)). ii)~kinematic effects, when
there are two maser clouds along the line of sight with different
velocities, such that a hyperfine transition in the foreground cloud
amplifies emission from a different hyperfine transition in the
background cloud\cite{walker:84}, iii)~population inversion of the
levels involved in maser action is preceded by collisional and radiative
de-excitation of higher torsion-rotation levels\cite{cragg:05}, with
rate coefficients that are hyperfine-state specific\cite{corey:83}. The
latter effect has been overlooked in current maser excitation
models\cite{cragg:05}, thus no quantitative information is available. To
obtain a qualitative understanding, we considered the relative
hyperfine-specific collisional and radiative rates within a
torsion-rotation transition.
We find for de-excitation
collisions of methanol with helium atoms (equation~(8)), and for radiative emission (equation~(12)), that the
hyperfine levels with the highest $F$ quantum number have the largest
relative de-excitation rate coefficients.

Up to now, methanol maser circular polarization observations have been
made for the 6.7 GHz ($5_{15}\ A_2 \to 6_{06} \
A_1$)\cite{vlemmings:11, surcis:12}, 44 GHz ($7_{07}\ A_2 \to 6_{16}\
A_1$)\cite{sarma:11,momjian:16} and 36 GHz ($4_{-1} \ E \to
3_{0} \ E$)\cite{sarma:09} torsion-rotation
transitions. As the magnetic characteristics of methanol were not known,
(hyperfine unspecific) estimates of the Zeeman parameters were used. In
the following, we will re-analyze some of the observations using our
calculated Zeeman parameters (Supplementary Table~1). We take into account that within
a torsion-rotation transition the various hyperfine transitions have
different Land\'{e} g-factors (Supplementary Tables~2-18) and that
the maser action can be hyperfine-state specific.

\begin{table}[t]
\caption{Zeeman splitting parameters $\alpha_Z$ (in Hz mG$^{-1}$) for the strongest
$\Delta F = \Delta J$ transitions of the investigated maser lines. The
ranges of the quantum number $F$ for states of $A$ symmetry are explained
in the caption of Figure~1.
Torsion-rotation states of $E$ symmetry state have nuclear spin $I$ = 0
and 1, so that for $J\geq 1$ there are hyperfine states with $F=J$ and
$F=J\pm 1$. Each torsion-rotation function of $E$ symmetry yields two
sets of hyperfine states of overall symmetry $A_1$ and
$A_2$\cite{lankhaar:16}. In this table, the hyperfine lines for each
$J\to J'$ torsion-rotation transition are indicated by their initial $F$
value, relative to the corresponding $J$ value. For $A$-symmetry a
single transition is associated with each $F=J\pm 2$ and two transitions
with each $F=J, J\pm 1$. For $E$-symmetry two transitions $A_1 \to A_2$
and $A_2 \to A_1$ are associated with each $F=J\pm 1$ and four
transitions with $F=J$.}
\centering
\begin{tabular}{l@{\extracolsep{0.6cm}}c@{\extracolsep{0.6cm}}c@{\extracolsep{0.15cm}}c@{\extracolsep{0.6cm}}c@{\extracolsep{0.15cm}}c@{\extracolsep{0.6cm}}c@{\extracolsep{0.15cm}}c@{\extracolsep{0.6cm}}c}
\hline \hline  & $F=J-2$ & \multicolumn{2}{c}{$F=J-1$} & \multicolumn{2}{c}{$F=J$} & \multicolumn{2}{c}{$F=J+1$} & $F=J+2 $ \\ \hline
$5_{15}\ A_2 \to 6_{06} \ A_1$ ($6.7$ GHz) & $-1.135$ & $-0.516$ & $-0.467$ &
$-0.127$ & $0.002$ & $0.224$ & $0.261$ & $0.472$ \\ $7_{07}\ A_2 \to
6_{16}\ A_1$ ($44$ GHz) & $-0.920$ & $-0.436$ & $-0.403$ & $-0.016$ & $-0.108$ &
$0.207$ & $0.203$ & $0.413$ \\ \multirow{2}{*}{$4_{-1} \ E \to 3_{0} \
E$ ($36$ GHz)} & & \multicolumn{2}{c}{$-0.704$} & $-0.075$ & $0.056$ &
\multicolumn{2}{c}{$0.424$} & \\ & & \multicolumn{2}{c}{$-0.729$} &
$-0.274$ & $0.174$ & \multicolumn{2}{c}{$0.486$} & \\ \hline \hline
\end{tabular} \end{table}

We begin with the circular-polarization observations of class II 6.7 GHz
methanol masers occurring in protostellar disks. We assume that the
transition with the largest Einstein coefficient for stimulated
emission, the $F = 3 \to 4$ transition (see Figure~1 and Supplementary
Table~3), will be favored and that the maser action is limited to this
transition. Then, the Zeeman-splitting coefficient $\alpha_Z$ (related
to the Land\'{e} g-factor $g_l$ as $\alpha_Z = \mu_N g_l$ with $\mu_N$
being the nuclear magneton) of the maser-transition will be $\alpha_Z =
-1.135$ Hz~mG$^{-1}$, which is 10 times larger than the value currently
used for magnetic field estimates\cite{vlemmings:11,jen:51}.
In the methanol maser
regions probed by these class II masers, with an H$_2$ number density of $n_{\rm
H_2}\approx 10^8$ cm$^{-3}$\cite{cragg:05}, application of our new
results to a large sample of maser
observations\cite{vlemmings:11}
indicates an average field strength of $\braket{|B|}\approx 12$~mG.
If, instead of by the $F=3 \to 4$ hyperfine transition, the polarization is caused
by a combination of hyperfine components or by any of the other components,
the derived magnetic field strength would be higher. Including all hyperfine
components would result in an average $\alpha_Z \approx 0.17$
Hz~mG$^{-1}$ and $\braket{|B|}\approx 80$ mG. This is significantly
larger than expected based on OH masers observed at similar
densities\cite{baudry:98,wright:04}.
The results based on the $F=3 \to 4$ transition are in good agreement with
OH-maser
polarization observations\cite{baudry:98,wright:04}, as well as with the extrapolated
magnetic field vs.\ density relation $B \propto n^{1/2}$~\cite{crutcher:99,vlemmings:08}.
This indicates, as
already suggested by linear polarization
studies\cite{vlemmings:10,surcis:12}, that methanol masers probe the
large scale magnetic field around massive protostars. Reversely,
extending the magnetic field vs.\ density relation by almost two orders
of magnitude in density provides important constraints on the theory of
massive star formation, as it implies that the magnetic energy density
remains important up to densities of $n_{\rm H_2}\approx
10^9$~cm$^{-3}$. The conclusions are also supported by a more specific
study of Cepheus A~HW2\cite{vlemmings:10} (Figure~3),
where our reinterpretation confirms a slightly supercritical maser
region, giving rise to magnetically regulated accretion towards the disc
of Cepheus~A~HW2.

Polarization observations of class I methanol masers in the outflows of
massive star-forming regions have been made for the 36~GHz and 44~GHz
torsion-rotation transitions. In both lines the individual hyperfine
transitions are clustered in two groups, which gives rise to a doublet
structure in the spectra\cite{lankhaar:16}. Considering kinematic
effects favoring one of the two peaks and selecting from this peak the
transition with the largest coefficient for stimulated emission, we
assume that the $F= 3 \to 2$ (36~GHz) and $F = 5\to 4$ (44~GHz)
hyperfine lines are favored and that the maser action is limited to
these transitions. Then, the Zeeman-splitting coefficients of the
maser-transitions will be $\alpha_Z = -0.704$ Hz~mG$^{-1}$ for the
36~GHz line and $\alpha_Z = -0.920$ Hz~mG$^{-1}$ for the 44 GHz line (Supplementary Tables~10 and 4).
The observed class I methanol masers are expected to occur in shocked
regions of the outflows at densities of an order of magnitude lower in
comparison to class II masers\cite{voronkov:06}. The
Zeeman splitting of the 36~GHz and 44~GHz lines was found to be on the
order of several tens of Hz\cite{sarma:09,sarma:11,momjian:16}. Using
our analysis, this would indicate magnetic field strengths of
$20-75$~mG. Since, class I masers are shock excited, shock compression is
expected to increase the magnetic field strength. For the outflow velocities in the
class I maser regions, a pre-shock magnetic field as observed in the OH
maser regions ($\approx 5$~mG) will be amplified to $>20$~mG, consistent
with the observations. We thus suggest that class I methanol maser
polarization observations provide important information on the shock
conditions of proto-stellar outflows.

Our results also suggest an explanation for a surprising feature
observed in both Class II (6.7 GHz) and Class I (44 GHz) masers.
Observations have shown reversals in the sign of polarization over areas
of small angular extent in the sky (6.7 GHz, see Figure~3 and 44 GHz, see Figure~2 in Ref.~\cite{momjian:16}).
Such reversals have usually been interpreted as a change in field
direction. However, reversals on au-scales would be surprising if one
considers the agreement between the fields probed by methanol masers and
dust emission\cite{dalolio:17}. A more plausible explanation favored by
our results is that in the masers with opposite signs of polarization,
the masing process itself is due to the dominance of different hyperfine
transitions. If we assume for the 6.7 GHz spectrum that for the
`recalcitrant' maser the $F = 7 \to 8$ hyperfine transition is favored
by kinematic effects, instead of the $F = 3 \to 4$ transition for the
other masers, we find a magnetic field comparable to the result from
other masers along the line of sight (Figure~3).
For the 44 GHz maser, if we assume the $F =
8 \to 7$ transition to be favored for the `recalcitrant' maser instead
of the $F = 5 \to 4$ transition for the other maser, we also get
Zeeman-splitting coefficients with opposite signs and we find similar magnetic fields of
$\approx 50$~mG from both masers composing the signal. We thus find that
an alternative preferred hyperfine transition in the maser action is
able to explain opposite circular polarization along the line of sight,
and obtain magnetic fields comparable with the results from other masers
that trace similar areas around the protostar.

Our model is also important for the study of methanol maser absorption in red-shifted
cosmological sources. Methanol's high sensitivity to variation of the
electron-to-proton mass ratio in the torsion-rotational structure is
enhanced by its torsional motion\cite{jansen:11a}. Extra-galactic
absorption measurements of the $3_{-1} \ E \to 2_{0}\ E$, 12.2 GHz
transition have been used to provide the strongest constraints on the
time variation of the electron-to-proton mass
ratio\cite{bagdonaite:13b,kanekar:15}. Recently,
measurements with a high spatial resolution have been able to
selectively observe methanol absorption in an extra-galactic cold
core\cite{marshall:16}. Hyperfine effects shift the center of
torsion-rotation lines, which is an effect not accounted for in the
current torsion-rotation fitting Hamiltonian\cite{xu:08} from which the
parameters are used in determination of the sensitivity
coefficients\cite{jansen:11a,jansen:11b}. Also, in cold extra-galactic
regions, temperature broadening effects are smaller than the hyperfine
splittings, which could be resolved with a sufficiently high spectral
resolution. Furthermore, in regions with strong magnetic fields
($>30$~mG), this structure will be affected by Zeeman effects. These
effects should be included in the error-accounting of the constraints
to the time variation of the electron-to-proton mass ratio.

Theoretical modeling of (non-linear) Zeeman effects for other molecular species such
as HCN and H$_2$CO can also be done according to the theory
presented here. The same care should be taken in the assessment of
Zeeman effects in radical species, where the magnetic field can mix
fine-structure states, which is analogous to hyperfine mixing. This will
be particularly important for CCS\cite{shinnaga:00, ramos:06}, for which
the Zeeman characteristics are still poorly known, but which will be one
of the prime molecules for Zeeman studies with the Square Kilometer
Array.

\section*{Methods}
We theoretically modeled the response of methanol to weak magnetic
fields by the addition of magnetic field (Zeeman) interactions to the
model for methanol's hyperfine structure from Ref.~\cite{lankhaar:16}.
Here, we will briefly revisit methanol's hyperfine structure and
describe the relevant Zeeman interactions. Next, we will detail the
computational methods used to obtain the molecule-specific coupling
parameters. Finally, we describe the methods used to compute the
magnetic field dependent spectrum of methanol.
\subsection*{Hyperfine structure}
The elucidation of methanol's hyperfine structure has been a challenging
problem. The CH$_3$-group in methanol can easily rotate with respect to
the OH-group, which leads to an extension of the usual rigid-rotor
hyperfine Hamiltonian with nuclear spin-torsion
interactions\cite{heuvel:73a}. In contrast with the nuclear
spin-rotation coupling parameters, for which the \textit{ab-initio}
calculated values have recently been experimentally confirmed\cite{belov:16},
the torsional hyperfine coupling
parameters cannot be obtained from quantum chemical \textit{ab initio}
calculations \cite{coudert:15, lankhaar:16}. Experiments probing the
hyperfine structure of methanol have proven difficult to interpret,
because the hyperfine transitions cannot be individually resolved.
Lankhaar \textit{et al.}\cite{lankhaar:16} revised the derivation of a
Hamiltonian which includes the torsional hyperfine interactions and
obtained the coupling parameters in this Hamiltonian from both
\textit{ab initio} calculations and experimental data\cite{heuvel:73a,
coudert:15}. The hyperfine spectra of methanol calculated from this
Hamiltonian agree well with the spectra observed for several
torsion-rotation states of both $A$- and $E$-symmetry. In our present
calculations of the Zeeman interactions of methanol in external
magnetic fields we start from this hyperfine Hamiltonian. For a detailed
description, see Ref.~\cite{lankhaar:16}.

\subsection*{Zeeman Hamiltonian}
Zeeman interactions are governed by the same magnetic moments that
determine the hyperfine structure, interacting with an external magnetic
field $\boldsymbol{B}$. For a closed-shell diamagnetic molecule as
methanol three contributions are important, from the overall rotation,
the internal rotation or torsion, and the nuclear spins. The most
abundant $^{12}$C and $^{16}$O nuclei have spin zero, so the nuclear
spin of methanol, CH$_3$OH, comes from the three protons in the CH$_3$
group and the proton in the OH group. As it was derived in Appendix A of
Ref.~\cite{lankhaar:16} for the corresponding hyperfine Hamiltonian, the
rotational Zeeman Hamiltonian
\begin{equation}
\hat{H}_{\mathrm{BR}} = - \frac{\mu_N}{\hbar} \boldsymbol{B} \cdot
\boldsymbol{g}(\gamma) \hat{\boldsymbol{J}}
+ \frac{\mu_N}{\hbar} f \boldsymbol{B} \cdot \boldsymbol{g}(\gamma) \boldsymbol{\lambda}
\left( \hat{p}_{\gamma} - \boldsymbol{\rho}\cdot \hat{\boldsymbol{J}}\right),
\label{eq:rotation}
\end{equation}
depends not only on the overall rotation angular momentum
$\hat{\boldsymbol{J}}$, but also on the torsional angular momentum
$\hat{p}_{\gamma}=(\hbar/i) \partial/\partial \gamma$. $\mu_N$ is the
nuclear magneton. The coupling tensor $\boldsymbol{g}$
has the same form as for semi-rigid molecules, but for molecules with
internal rotation it depends on the torsional angle $\gamma$. The unit
vector $\boldsymbol{\lambda}$ describes the direction of the internal
rotation axis in the principal axes frame of the molecule and
$\boldsymbol{\rho}=\boldsymbol{I}^{-1}\boldsymbol{I}^{\hbox{CH$_3$}}
\boldsymbol{\lambda}$, where $\boldsymbol{I}$ is the total inertia
tensor and $\boldsymbol{I}^{\hbox{CH$_3$}}$ the inertia tensor of the
CH$_3$ group. The dimensionless factor $f$ depends on the ratio of the
moments of inertia of the OH frame and the rotating CH$_3$ top about the
torsional axis\cite{lankhaar:16}.

In addition, we must account for torsional Zeeman effects. Similarly to
the torsional hyperfine Hamiltonian $\hat{H}_{\mathrm{ST}}$ in
Ref.~\cite{lankhaar:16}, the torsional Zeeman Hamiltonian
\begin{equation}
\hat{H}_{\mathrm{BT}} = -\frac{\mu_N}{\hbar} f \boldsymbol{B} \cdot
\boldsymbol{b}(\gamma) \left(\hat{p}_{\gamma} - \boldsymbol{\rho}\cdot \hat{\boldsymbol{J}} \right),
\label{eq:torsion}
\end{equation}
with the coupling vector $\boldsymbol{b}(\gamma)$, not only contains the
torsional angular momentum operator $\hat{p}_{\gamma}$, but also the
total angular momentum $\hat{\boldsymbol{J}}$. By absorbing the second
term of $\hat{H}_{\mathrm{BR}}$ in equation~(\ref{eq:rotation}) into
equation~(\ref{eq:torsion}), the remaining rotational Zeeman Hamiltonian
obtains the usual form it has for a semi-rigid molecule, and we obtain
an effective torsional Hamiltonian $\hat{H}_{\mathrm{BT}}$ with $\boldsymbol{b}(\gamma)$
replaced by
\begin{equation}
\boldsymbol{b}'(\gamma) = \boldsymbol{b}(\gamma) - \boldsymbol{g}(\gamma)\boldsymbol{\lambda},
\end{equation}
Finally, the intrinsic magnetic moments of the protons $K=1,2,3$ in the
CH$_3$ group and proton $K=4$ in the OH group interact with the magnetic
field
\begin{equation}
\hat{H}_{\mathrm{BS}} = -\frac{\mu_N}{\hbar} g_p \sum_K \boldsymbol{B} \cdot
\hat{\boldsymbol{I}}_K,
\end{equation}
where $g_p$ is the proton g-factor. The total Zeeman Hamiltonian is a sum of
the rotational, torsional, and nuclear spin Zeeman terms
\begin{equation}
\hat{H}_{\mathrm{Zeeman}} = \hat{H}_{\mathrm{BR}} + \hat{H}_{\mathrm{BT}} + \hat{H}_{\mathrm{BS}} .
\label{eq:zeeman_ham}
\end{equation}

\subsection*{Coupling tensors}
The response of methanol to magnetic fields is theoretically modeled by
including the coupling of the relevant angular momenta ---the rotational
angular momentum, the torsional momentum, and the nuclear spin angular
momentum--- to the magnetic field vector. The couplings between these
angular momentum operators and the magnetic field involve a rank-2
coupling tensor and a coupling vector. This coupling tensor and vector
are specific for methanol. The derivation of the hyperfine coupling
tensors is given in Ref.~\cite{lankhaar:16}, and of the Zeeman coupling
tensors in the Supplementary Information. This subsection
describes the methods used to evaluate all coupling parameters.
\subsubsection*{Rotational g-tensor}
Rotational Zeeman effects are represented by the molecule-specific
g-tensor, which for rigid non-paramagnetic molecules has
been extensively studied experimentally for its valuable information on
the electronic structure\cite{eshbach:52, flygare:71, sutter:76,
flygare:77}. Nowadays, quantum chemical
calculations\cite{gauss:96a,lutnaes:09} are able to reproduce these
experiments with high accuracy. The rotational g-tensor
$\boldsymbol{g}(\gamma)$ can be obtained from \textit{ab initio}
electronic structure calculations with the program package
CFOUR\cite{cfour}. We carried out calculations with CFOUR at the
coupled-cluster level of theory including single and double excitations
with perturbative addition of the triples contribution [CCSD(T)], in an
augmented triple-zeta correlation-consistent (aug-cc-pVTZ) basis
set\cite{dunning:89}. The geometry of methanol was optimized at this
level, which yields bond lengths OH, CO, and CH of 0.956, 1.427, and
1.096~\AA, respectively, bond angles COH and OCH of 108.87$^\circ$
and 109.91$^\circ$, and a torsional HOCH angle of 180$^\circ$. The
electronic contributions to $\boldsymbol{g}(\gamma)$ were
calculated at the same level of theory for 13 equidistant values of the
torsional angle $\gamma$ by keeping the HOC fragment fixed and rotating
the CH$_3$ group over these angles about the OC bond axis. The nuclear
contribution to the tensor $\boldsymbol{g} (\gamma)$ was also given by
CFOUR, but was also calculated directly from the nuclear coordinates.
Because of methanol's symmetry, we could fit our \textit{ab initio}
calculated values for the rotational
$\boldsymbol{g} (\gamma)$-tensor elements to $\sum_n a_{3n} \cos(3n \gamma)$ or
$\sum_n a_{3n} \sin(3n \gamma)$ functions of the internal rotation angle.
The expansion coefficients, $a_{3n}$, are listed in Supplementary Table~1.

\subsubsection*{Torsional b-vector}
Torsional Zeeman interactions are represented by the molecule-specific
b-vector (Supplementary equations~(16) and~(21)). The calculation of
the electronic contribution to the b-vector has not been implemented in
the available quantum-chemical program packages. In order to estimate
the torsional Zeeman effects in methanol, we compare its internally
rotating CH$_3$-group to the CH$_3$-groups of nitromethane and
methyl-boron-difluoride, of which the torsional Zeeman effect has been
investigated experimentally\cite{engelbrecht:73, engelbrecht:75}.

The torsional Zeeman coupling vectors of nitromethane and
methyl-boron-difluoride were determined to be $\boldsymbol{b} =
g_{\gamma} \boldsymbol{\lambda}$, with $g_{\gamma}=0.347$ and $0.3415$,
respectively. In these molecules the unit vector $\boldsymbol{\lambda}$
that defines the direction of the internal rotation axis lies along the
main principal axis. The small difference in the g-values of these two
molecules was explained by the electron drainage from the CH$_3$-groups
by the attached functional group (see Supplementary Information). This
electron drainage can be estimated from the partial atomic charges given
by a Mulliken population analysis\cite{mulliken:55}. We calculated
Mulliken populations of the CH$_3$-groups of nitromethane and
methyl-boron-difluoride at the CCSD(T) level in an aug-cc-pVDZ basis, at
their \textit{ab initio} optimized geometries, and found that
$P_{\mathrm{CH}_3(\mathrm{-NO}_2)} = 8.714$ and
$P_{\mathrm{CH}_3(\mathrm{-BF}_2)} = 9.232$. The Mulliken population of
the CH$_3$-group in methanol was computed at the same level and found
to be $P_{\mathrm{CH}_3(\mathrm{-OH})} = 8.737$. Then, we obtained
the b-vector of methanol by interpolation as
$\boldsymbol{b} = g_{\gamma} \boldsymbol{\lambda}$, with $g_{\gamma} =
0.3468$, see Supplementary Table~1. In this estimate of
$\boldsymbol{b}$, we have assumed that it is independent of the
internal rotation angle $\gamma$ and is parallel to
$\boldsymbol{\lambda}$. The latter assumption holds only when
$\boldsymbol{\lambda}$ is directed along one of the pricipal axes, which
is almost the case for methanol\cite{xu:08,lankhaar:16}.

\subsection*{Matrix elements and spectrum}
With the knowledge of the coupling tensors $\boldsymbol{g}(\gamma)$ and
$\boldsymbol{b}(\gamma)$ in the Zeeman Hamiltonian of
equation~(\ref{eq:zeeman_ham}) and the use of the hyperfine Hamiltonian
from Ref.~\cite{lankhaar:16} we computed the magnetic field dependence
of the hyperfine levels. The total Hamiltonian is diagonalized in the
basis $\ket{\{(I_{123}, I_4) I, J\} F M_F}$ obtained by coupling the
eigenfunctions of the torsion-rotation Hamiltonian\cite{xu:08} with the
nuclear spin functions $\ket{(I_{123}, I_4) I}$ of the appropriate
symmetry, defined in Sec.~IIC of Ref.~\cite{lankhaar:16}. The hyperfine
interactions couple the total nuclear spin $I$ with the torsion-rotation
angular momentum $J$ to a total angular momentum $F$, with projection
$M_F$ on the space-fixed $z$-axis chosen along the magnetic field
direction $\boldsymbol{B}$. When the external magnetic field is included,
only $M_F$ remains a good quantum number. Hyperfine states with
different $F$ may get mixed, which happens to a substantial extent when
there is a small energy gap between the hyperfine levels.

The torsion-rotation wave functions\cite{xu:08} have quantum numbers
$v_\tau,\ J,\ K_a$ and symmetry $A$ or $E$. For symmetries $A$ and $E$,
the nuclear spin basis has $I_{123}=3/2$ and 1/2, respectively, see
Sec.~IIC in Ref.~\cite{lankhaar:16}. The energy gaps between
torsion-rotation states are typically on the order of a few GHz, while
the hyperfine and Zeeman interactions in methanol (for fields $B<10$~G)
amount to about 10~kHz. Hence, we may restrict our basis to a single
value of $v_\tau$ and $J$ and derive an effective Zeeman Hamiltonian
of which the matrix elements are more easily evalutated (Supplementary Information).

The matrix of the total Zeeman plus hyperfine Hamiltonian over the
hyperfine basis with quantum numbers $F, M_F$\cite{lankhaar:16} is
evaluated and diagonalized. This yields the splitting of the hyperfine
levels into $2F+1$ sublevels, with $M_F$ as the only good quantum number.
Intensities and Einstein A-coefficients for transitions between the
individual hyperfine levels are obtained with the procedures described
in Ref.~\cite{lankhaar:16}.

\subsection*{Hyperfine-state resolved de-excitation}
It is emphasized in the letter that the Land\'{e} g-factors of the
different hyperfine components of each torsion-rotation transition vary
over a large range of values. It is therefore important to know the
populations of the individual hyperfine levels of the torsion-rotation
states involved in the methanol maser action. This maser action is
preceded by collisional and radiative de-excitation of higher
torsion-rotation levels. Here, we derive formulas to estimate relative
hyperfine-state-specific collisional and radiational de-excitation rate
coefficients.

\subsubsection*{Hyperfine-state resolved collisional rate coefficients}
Hyperfine splittings are negligible with respect to the collision
energy, so to an excellent approximation the collision dynamics depends
only on the scattering conditions and on the torsion-rotation structure
of the molecule, and is not affected by hyperfine effects\cite{corey:83,
alexander:85}. As a consequence, obtaining hyperfine-state-specific
$F\to F'$ transition rate coefficients from the usual rate coefficients
for rotationally inelastic $J\to J'$ collisions requires only the use of
an appropriate basis in which the angular momentum $J$ is coupled
with the total nuclear spin $I$ to total angular momentum
$F$\cite{corey:83,neufeld:94}.

Davis\cite{davis:91} analyzed the collision dynamics of a structureless
atom (such as helium) and a molecule with one internal rotation (as
methanol), using a simple torsion-rotation model and neglecting the
molecule's vibrational modes. The presence of the torsional modes leads
to additional inelastic scattering processes, but the angular momentum
algebra in the expressions for the scattering amplitudes and cross
sections are similar for collisions with methanol and with diatomic
molecules. The methanol symmetric rotor quantum number $K_a$
and torsional quantum numbers $v_{\tau}$ and $\sigma$ only play a
spectator role with regard to the angular momentum
(re-)coupling\cite{davis:91}.

To perform coupled-channel calculations for collisions of methanol with
helium in a hyperfine basis would exceed the scope of this letter.
Rather, we want to get an idea of the general trends of the hyperfine
inelastic scattering rates with respect to the corresponding rotational
rates. To this end, we use a formalism that relates the torsion-rotation
inelastic scattering cross sections to hyperfine-state-specific cross
sections. This is most conveniently done by representing the cross
sections in terms of tensor opacities\cite{alexander:85, davis:91}
\begin{equation}
\label{eq:inelasrot}
\sigma_{(K_a \sigma v_{\tau}) J \to (K'_a \sigma' v_{\tau}') J'} =
\frac{\pi}{[J] k_{(K_a \sigma v_{\tau})J}^2 } \sum_L P^{(L)}
((K_a \sigma v_{\tau}) J \to (K'_a \sigma' v_{\tau}') J'),
\end{equation}
where $k_{(K_a \sigma v_{\tau})J}$ is the wave vector dependent on the
collision energy and the torsion-rotation energy of the initial state,
$L$ is the rank of the tensor opacity $P^{(L)}$, and the square bracket
notation designates $[J] = 2J+1$. Recoupling the
nuclear-spin free opacity to the hyperfine-state-specific opacity
yields\cite{alexander:85}
\begin{equation}
P^{(L)} ((K_a \sigma v_{\tau})[JI]F \to (K_a' \sigma' v_{\tau}')[J'I]F')
= [F][F']\begin{Bmatrix} J & J' & L \\ F' & F & I \end{Bmatrix}^2 P^{(L)} ((K_a \sigma v_{\tau})J \to (K_a' \sigma' v_{\tau}')J').
\label{eq:opac}
\end{equation}
The expression in curly brackets is a $6j$-symbol.
Substitution of this result into equation~(\ref{eq:inelasrot}), yields
the hyperfine-state-specific collisional cross sections
\begin{eqnarray}
\label{eq:inelashyp}
\sigma_{(K_a \sigma v_{\tau}) [JI]F \to (K_a' \sigma' v_{\tau}') (J'I)F'}
& = & \frac{\pi}{k_{(K_a \sigma v_{\tau})J}^2[F]} \sum_L P^{(L)} ((K_a \sigma v_{\tau})[JI]f \to (K_a' \sigma' v_{\tau}')[J'I]F')
\\
& = & \frac{\pi [F']}{k_{(K_a \sigma v_{\tau})J}^2} \sum_L
\begin{Bmatrix} J & J' & L \\ F' & F & I \end{Bmatrix}^2 P^{(L)} ((K_a \sigma v_{\tau})J \to (K_a' \sigma' v_{\tau}')J').
\nonumber
\end{eqnarray}

The triangular condition imposes the constraint $|J-J'| \le L \le J+J'$.
Each of these values of $L$ contributes to the
total hyperfine-state-specific collisional rates. Except for very low
collision energies where resonances may occur, which are not important
at the methanol maser conditions,
the largest contribution comes from $L = |\Delta J|$. Considering this
contribution only would directly relate the hyperfine-state-specific
rate with the rotational rate\cite{neufeld:94}
\begin{equation}
\label{eq:hyperrate_approx}
\sigma_{(K_a \sigma v_{\tau}) [JI]F \to (K_a' \sigma' v_{\tau}') [J'I]F'} = [J][F']
\begin{Bmatrix} J'& J & |\Delta J| \\ F & F' & I \end{Bmatrix}^2  \sigma_{(K_a \sigma v_{\tau}) J \to (K_a' \sigma' v_{\tau}') J'}.
\end{equation}
To analyze which final hyperfine levels, $F'$, are mostly populated by
(de-)excitation, we must sum equation~(\ref{eq:hyperrate_approx}) over all
initial hyperfine states $F$.
Then, we find for example for the collisional de-excitation $J=6
\to J'=5$ that the $F'=J'+2$ state has a 14\% higher propensity
to be populated than the $F' = J'-2$ state. The other hyperfine states
with $F'=J'$ and $J'\pm 1$ are linear combinations of two nuclear spin
states, and have population propensities lower than the $F'=J'+2$ state
and higher than $F' = J'-2$ state.

Actually, one can also analyze the ratio between hyperfine rates
including all $L$ channels with equation~(\ref{eq:inelashyp}), and find that hyperfine-state-specific collision
propensities for the $F'=J'+2$ state are even higher for the
other $L$ channels. We can therefore say that the hyperfine collisional
de-excitation propensity is over $14\%$ higher for the $F'=J'+2$ state than
for the $F'=J'-2$ state. Hyperfine state specific collision rate
propensities of intermediate $F'$ states are somewhere between the two
extremes. More in general, we find consistently that in collisional
(de-)excitation, for $\Delta J < 0$ transitions, high $F'$ states have a
higher propensity for every $L$ channel. In $\Delta J = 0 $ transitions,
there is no hyperfine preference, and in $\Delta J > 0 $ transitions,
low $F'$ hyperfine states have the highest probability to be populated.

\subsubsection*{Hyperfine-state resolved radiative rates}
To investigate which hyperfine transitions are favored by
radiative de-excitation, we will
recouple the line strength of a torsion-rotation transition to a
hyperfine basis. From the line strength, one can easily compute the
Einstein coefficient.

The line strength of a transition between torsion-rotation states $(K_a \sigma v_{\tau})J$
and $(K_a' \sigma' v_{\tau}')J'$ is
\begin{equation}
S_{(K_a \sigma v_{\tau})J \to (K_a' \sigma' v_{\tau}')J' } = \left| \sum_{M M'}
 \braket{(K_a \sigma v_{\tau})J M| \boldsymbol{d} | (K_a' \sigma' v_{\tau}')J' M' } \right|^2,
\end{equation}
where $\boldsymbol{d}$ stands for the total dipole moment vector of all
charged particles, electrons and nuclei.
In recoupling the rotational line strength to hyperfine-state specific line strengths,
we recognize that the line strength transforms as a rank-1 tensor
opacity, and we use equation~(\ref{eq:opac}) to obtain
\begin{equation}
S_{(K_a \sigma v_{\tau})[JI]F \to (K_a' \sigma' v_{\tau}')[J'I]F'} =
[F] [F'] \begin{Bmatrix} J'& J & 1 \\ F & F' & I \end{Bmatrix}^2
 S_{(K_a \sigma v_{\tau})J \to (K_a' \sigma' v_{\tau}')J'}.
\end{equation}
The Einstein coefficient is related to the line strength by
\begin{equation}
A_{(K_a \sigma v_{\tau})[JI]F \to (K_a' \sigma' v_{\tau}')[J'I]F'} = \frac{2 \omega^3}{3 \epsilon_0 h c^3 [F]} S_{(K_a \sigma v_{\tau})[JI]F \to (K_a' \sigma' v_{\tau}')[J'I]F'},
\end{equation}
where $\epsilon_0$, $h$ and $c$ are the vacuum permitivity, Planck's
constant, and the speed of light, respectively. $\omega$ is the
transition frequency, which to a very good approximation does not
depend on the hyperfine splittings of the rotational states.

To analyze which final hyperfine levels $F'$ are mostly populated by
(de-)excitation, we must sum equation~(\ref{eq:hyperrate_approx}) over all
initial hyperfine states $F$: $A_{(K_a \sigma v_{\tau})J \to (K_a' \sigma' v_{\tau}')[J'I]F'}$.
Then, we find for a $J=6 \to J'=5$ emission, for example, that the
ratio between the Einstein coefficients for de-excitation to the
final hyperfine states with $F'=J'+2$  and $F' = J'-2$ is
\begin{equation}
\frac{A_{(K_a \sigma v_{\tau})6 \to (K_a' \sigma' v_{\tau}')[5I]7}}{A_{(K_a \sigma v_{\tau})6 \to (K_a' \sigma' v_{\tau}')[5I]3}} = 1.14 \ .
\label{eq:ratratio}
\end{equation}
The intermediate hyperfine states with $F'=J'$ and $J'\pm 1$ are linear
combinations of two nuclear spin states, and have population
propensities lower than the $F'=J'+2$ state and higher than the $F' =
J'-2$ state. The ratio calculated in equation~(\ref{eq:ratratio}) increases
exponentially for lower $J$. For rotational states with higher $J$ it
decreases to 1. 

\section*{Data availability}
The data that support the plots within this paper and other findings of this study are available from the corresponding author upon reasonable request.

\section*{Correspondence}
To whom correspondence should be addressed: Boy Lankhaar (boy.lankhaar@chalmers.se)

\section*{Acknowledgements}
Support for this work was provided by the the Swedish Research Council (VR), and by the European Research Council under the European Union's Seventh Framework Programme (FP7/2007-2013), through the ERC consolidator grant agreement nr. 614264.

\section*{Author contributions}
B.L., A.vd.A. and W.V. wrote the paper. B.L.,
A.vd.A. and G.C.G. modeled the Zeeman
effect in methanol. B.L. and W.V. performed
the analysis of the astrophysical maser
spectra, based on methanol's Zeeman model.
W.V., H.J.vd.L. and G.S. provided
expertise on maser polarization in
astrophysics and initiated the project.
All authors discussed the results and
commented on the manuscript.

\section*{Competing financial interests}
The authors declare no competing financial interests.


\begin{thebibliography}{10}
\expandafter\ifx\csname url\endcsname\relax
  \def\url#1{\texttt{#1}}\fi
\expandafter\ifx\csname urlprefix\endcsname\relax\def\urlprefix{URL }\fi
\providecommand{\bibinfo}[2]{#2}
\providecommand{\eprint}[2][]{\url{#2}}

\bibitem{crutcher:12}
\bibinfo{author}{Crutcher, R.~M.}
\newblock \bibinfo{title}{Magnetic fields in molecular clouds}.
\newblock \emph{\bibinfo{journal}{Annu. Rev. Astron. Astrophys.}}
  \textbf{\bibinfo{volume}{50}}, \bibinfo{pages}{29--63}
  (\bibinfo{year}{2012}).

\bibitem{vlemmings:11}
\bibinfo{author}{{Vlemmings, W. H. T.}}, \bibinfo{author}{{Torres, R. M.}} \&
  \bibinfo{author}{{Dodson, R.}}
\newblock \bibinfo{title}{Zeeman splitting of 6.7 {GH}z methanol masers}.
\newblock \emph{\bibinfo{journal}{Astron. Astroph.}}
  \textbf{\bibinfo{volume}{529}}, \bibinfo{pages}{A95} (\bibinfo{year}{2011}).

\bibitem{sarma:09}
\bibinfo{author}{Sarma, A.} \& \bibinfo{author}{Momjian, E.}
\newblock \bibinfo{title}{Detection of the {Z}eeman effect in the 36 {GHz}
  {Class} {I} {CH}$_3${OH} maser line with the {EVLA}}.
\newblock \emph{\bibinfo{journal}{Astrophys. J. Lett.}}
  \textbf{\bibinfo{volume}{705}}, \bibinfo{pages}{L176} (\bibinfo{year}{2009}).

\bibitem{sarma:11}
\bibinfo{author}{Sarma, A.} \& \bibinfo{author}{Momjian, E.}
\newblock \bibinfo{title}{Discovery of the {Z}eeman effect in the 44 {GHz}
  {Class} {I} methanol ({CH}$_3${OH}) maser line}.
\newblock \emph{\bibinfo{journal}{Astrophys. J. Lett.}}
  \textbf{\bibinfo{volume}{730}}, \bibinfo{pages}{L5} (\bibinfo{year}{2011}).

\bibitem{jen:51}
\bibinfo{author}{Jen, C.~K.}
\newblock \bibinfo{title}{Rotational magnetic moments in polyatomic molecules}.
\newblock \emph{\bibinfo{journal}{Phys. Rev.}} \textbf{\bibinfo{volume}{81}},
  \bibinfo{pages}{197} (\bibinfo{year}{1951}).

\bibitem{lankhaar:16}
\bibinfo{author}{Lankhaar, B.}, \bibinfo{author}{Groenenboom, G.~C.} \&
  \bibinfo{author}{van~der Avoird, A.}
\newblock \bibinfo{title}{Hyperfine interactions and internal rotation in
  methanol}.
\newblock \emph{\bibinfo{journal}{J. Chem. Phys.}}
  \textbf{\bibinfo{volume}{145}}, \bibinfo{pages}{244301}
  (\bibinfo{year}{2016}).

\bibitem{baudry:98}
\bibinfo{author}{Baudry, A.} \& \bibinfo{author}{Diamond, P.}
\newblock \bibinfo{title}{{VLBA} polarization observations of the {J}= 7/2,
  13.44 {GHz} {OH} maser in {W3 (OH)}}.
\newblock \emph{\bibinfo{journal}{Astron. Astroph.}}
  \textbf{\bibinfo{volume}{331}}, \bibinfo{pages}{697--708}
  (\bibinfo{year}{1998}).

\bibitem{wright:04}
\bibinfo{author}{Wright, M.~M.}, \bibinfo{author}{Gray, M.~D.} \&
  \bibinfo{author}{Diamond, P.~J.}
\newblock \bibinfo{title}{The {OH} ground-state masers in {W3(OH)} {– II.}
  {P}olarization and multifrequency results}.
\newblock \emph{\bibinfo{journal}{Mon. Not. R. Astron. Soc.}}
  \textbf{\bibinfo{volume}{350}}, \bibinfo{pages}{1272--1287}
  (\bibinfo{year}{2004}).

\bibitem{bagdonaite:13b}
\bibinfo{author}{Bagdonaite, J.} \emph{et~al.}
\newblock \bibinfo{title}{A stringent limit on a drifting proton-to-electron
  mass ratio from alcohol in the early universe}.
\newblock \emph{\bibinfo{journal}{Science}} \textbf{\bibinfo{volume}{339}},
  \bibinfo{pages}{46--48} (\bibinfo{year}{2013}).

\bibitem{kanekar:15}
\bibinfo{author}{Kanekar, N.} \emph{et~al.}
\newblock \bibinfo{title}{Constraints on changes in the proton-electron mass
  ratio using methanol lines}.
\newblock \emph{\bibinfo{journal}{Mon. Not. R. Astron. Soc.}}
  \textbf{\bibinfo{volume}{448}}, \bibinfo{pages}{L104--L108}
  (\bibinfo{year}{2015}).

\bibitem{jansen:11a}
\bibinfo{author}{Jansen, P.}, \bibinfo{author}{Xu, L.-H.},
  \bibinfo{author}{Kleiner, I.}, \bibinfo{author}{Ubachs, W.} \&
  \bibinfo{author}{Bethlem, H.~L.}
\newblock \bibinfo{title}{Methanol as a sensitive probe for spatial and
  temporal variations of the proton-to-electron mass ratio}.
\newblock \emph{\bibinfo{journal}{Phys. Rev. Lett.}}
  \textbf{\bibinfo{volume}{106}}, \bibinfo{pages}{100801}
  (\bibinfo{year}{2011}).

\bibitem{dapra:17}
\bibinfo{author}{Dapr{\`a}, M.} \emph{et~al.}
\newblock \bibinfo{title}{Testing the variability of the proton-to-electron
  mass ratio from observations of methanol in the dark cloud core l1498}.
\newblock \emph{\bibinfo{journal}{Mon. Not. R. Astron. Soc.}}
  \bibinfo{pages}{stx2308} (\bibinfo{year}{2017}).

\bibitem{engelbrecht:75}
\bibinfo{author}{Engelbrecht, L.}
\newblock \emph{\bibinfo{title}{Der Rotations-{Z}eemaneffekt bei Molek{\"u}len
  mit schwach behinderter interner Rotation}}.
\newblock Ph.D. thesis, \bibinfo{school}{University of Kiel}
  (\bibinfo{year}{1975}).

\bibitem{deguchi:90}
\bibinfo{author}{Deguchi, S.} \& \bibinfo{author}{Watson, W.~D.}
\newblock \bibinfo{title}{Linearly polarized radiation from astrophysical
  masers due to magnetic fields when the rate for stimulated emission exceeds
  the {Z}eeman frequency}.
\newblock \emph{\bibinfo{journal}{Astrophys. J.}}
  \textbf{\bibinfo{volume}{354}}, \bibinfo{pages}{649--659}
  (\bibinfo{year}{1990}).

\bibitem{vlemmings:06}
\bibinfo{author}{{Vlemmings, W. H. T.}}, \bibinfo{author}{{Diamond, P. J.}},
  \bibinfo{author}{{van Langevelde, H. J.}} \& \bibinfo{author}{{Torrelles, J.
  M.}}
\newblock \bibinfo{title}{The magnetic field in the star-forming region
  {Cepheus A}. from {H}$\mathsf{_2}${O} maser polarization observations}.
\newblock \emph{\bibinfo{journal}{Astron. Astrophys.}}
  \textbf{\bibinfo{volume}{448}}, \bibinfo{pages}{597--611}
  (\bibinfo{year}{2006}).

\bibitem{walker:84}
\bibinfo{author}{Walker, R.}
\newblock \bibinfo{title}{{H}$_2${O} in {W49N}. {II}-{S}tatistical studies of
  hyperfine structure, clustering, and velocity distributions}.
\newblock \emph{\bibinfo{journal}{Astrophys. J.}}
  \textbf{\bibinfo{volume}{280}}, \bibinfo{pages}{618--628}
  (\bibinfo{year}{1984}).

\bibitem{cragg:05}
\bibinfo{author}{Cragg, D.}, \bibinfo{author}{Sobolev, A.} \&
  \bibinfo{author}{Godfrey, P.}
\newblock \bibinfo{title}{Models of class {II} methanol masers based on
  improved molecular data}.
\newblock \emph{\bibinfo{journal}{Mon. Not. R. Astron. Soc.}}
  \textbf{\bibinfo{volume}{360}}, \bibinfo{pages}{533--545}
  (\bibinfo{year}{2005}).

\bibitem{corey:83}
\bibinfo{author}{Corey, G.} \& \bibinfo{author}{McCourt, F.~R.}
\newblock \bibinfo{title}{Inelastic differential and integral cross sections
  for $^{2S+ 1}\sigma$ linear molecule-$^1${S} atom scattering: the use of
  {H}und's case b representation}.
\newblock \emph{\bibinfo{journal}{J. Phys. Chem.}}
  \textbf{\bibinfo{volume}{87}}, \bibinfo{pages}{2723--2730}
  (\bibinfo{year}{1983}).

\bibitem{surcis:12}
\bibinfo{author}{Surcis, G.}, \bibinfo{author}{Vlemmings, W. H.~T.},
  \bibinfo{author}{van Langevelde, H.~J.} \&
  \bibinfo{author}{Hutawarakorn~Kramer, B.}
\newblock \bibinfo{title}{{EVN} observations of 6.7 {GHz} methanol maser
  polarization in massive star-forming regions}.
\newblock \emph{\bibinfo{journal}{Astron. Astrophys.}}
  \textbf{\bibinfo{volume}{541}}, \bibinfo{pages}{A47} (\bibinfo{year}{2012}).

\bibitem{momjian:16}
\bibinfo{author}{Momjian, E.} \& \bibinfo{author}{Sarma, A.}
\newblock \bibinfo{title}{The {Z}eeman effect in the 44 {GHz} class {I}
  methanol maser line toward {DR21} {(OH)}}.
\newblock \emph{\bibinfo{journal}{Astrophys. J.}}
  \textbf{\bibinfo{volume}{834}}, \bibinfo{pages}{168} (\bibinfo{year}{2017}).

\bibitem{crutcher:99}
\bibinfo{author}{Crutcher, R.~M.}
\newblock \bibinfo{title}{Magnetic fields in molecular clouds: observations
  confront theory}.
\newblock \emph{\bibinfo{journal}{Astrophys. J.}}
  \textbf{\bibinfo{volume}{520}}, \bibinfo{pages}{706} (\bibinfo{year}{1999}).

\bibitem{vlemmings:08}
\bibinfo{author}{{W. H. T. Vlemmings}}.
\newblock \bibinfo{title}{A new probe of magnetic fields during high-mass star
  formation}.
\newblock \emph{\bibinfo{journal}{Astron. Astroph.}}
  \textbf{\bibinfo{volume}{484}}, \bibinfo{pages}{773--781}
  (\bibinfo{year}{2008}).

\bibitem{vlemmings:10}
\bibinfo{author}{Vlemmings, W.}, \bibinfo{author}{Surcis, G.},
  \bibinfo{author}{Torstensson, K.} \& \bibinfo{author}{Van~Langevelde, H.}
\newblock \bibinfo{title}{Magnetic field regulated infall on the disc around
  the massive protostar {Cepheus A HW2}}.
\newblock \emph{\bibinfo{journal}{Mon. Not. R. Astron. Soc.}}
  \textbf{\bibinfo{volume}{404}}, \bibinfo{pages}{134--143}
  (\bibinfo{year}{2010}).

\bibitem{voronkov:06}
\bibinfo{author}{Voronkov, M.~A.} \emph{et~al.}
\newblock \bibinfo{title}{Class i methanol masers in the outflow of iras 16
  547--4247}.
\newblock \emph{\bibinfo{journal}{Mon. Not. R. Astron. Soc.}}
  \textbf{\bibinfo{volume}{373}}, \bibinfo{pages}{411--424}
  (\bibinfo{year}{2006}).

\bibitem{dalolio:17}
\bibinfo{author}{{Dall'Olio}, D.} \emph{et~al.}
\newblock \bibinfo{title}{{Methanol masers reveal the magnetic field of the
  high-mass protostar IRAS 18089-1732}}.
\newblock \emph{\bibinfo{journal}{ArXiv e-prints}}  (\bibinfo{year}{2017}).
\newblock \eprint{1708.02961}.

\bibitem{marshall:16}
\bibinfo{author}{Marshall, M.~A.} \emph{et~al.}
\newblock \bibinfo{title}{Methanol absorption in {PKS B1830-211} at
  milliarcsecond scales}.
\newblock \emph{\bibinfo{journal}{Mon. Not. R. Astron. Soc.}}
  \textbf{\bibinfo{volume}{466}}, \bibinfo{pages}{2450} (\bibinfo{year}{2017}).

\bibitem{xu:08}
\bibinfo{author}{Xu, L.-H.} \emph{et~al.}
\newblock \bibinfo{title}{Torsion-rotation global analysis of the first three
  torsional states (v$_t$ = 0, 1, 2) and terahertz database for methanol}.
\newblock \emph{\bibinfo{journal}{J. Mol. Spectr.}}
  \textbf{\bibinfo{volume}{251}}, \bibinfo{pages}{305--313}
  (\bibinfo{year}{2008}).

\bibitem{jansen:11b}
\bibinfo{author}{Jansen, P.}, \bibinfo{author}{Kleiner, I.},
  \bibinfo{author}{Xu, L.-H.}, \bibinfo{author}{Ubachs, W.} \&
  \bibinfo{author}{Bethlem, H.~L.}
\newblock \bibinfo{title}{Sensitivity of transitions in internal rotor
  molecules to a possible variation of the proton-to-electron mass ratio}.
\newblock \emph{\bibinfo{journal}{Phys. Rev. A}} \textbf{\bibinfo{volume}{84}},
  \bibinfo{pages}{062505} (\bibinfo{year}{2011}).

\bibitem{shinnaga:00}
\bibinfo{author}{Shinnaga, H.} \& \bibinfo{author}{Yamamoto, S.}
\newblock \bibinfo{title}{Zeeman effect on the rotational levels of {CCS} and
  {SO} in the $^3\sigma$-ground state}.
\newblock \emph{\bibinfo{journal}{Astrophys. J.}}
  \textbf{\bibinfo{volume}{544}}, \bibinfo{pages}{330} (\bibinfo{year}{2000}).

\bibitem{ramos:06}
\bibinfo{author}{Ramos, A.~A.} \& \bibinfo{author}{Bueno, J.~T.}
\newblock \bibinfo{title}{Theory and modeling of the {Zeeman} and
  {Paschen}-{Back} effects in molecular lines}.
\newblock \emph{\bibinfo{journal}{Astrophys. J.}}
  \textbf{\bibinfo{volume}{636}}, \bibinfo{pages}{548} (\bibinfo{year}{2006}).

\bibitem{heuvel:73a}
\bibinfo{author}{Heuvel, J.} \& \bibinfo{author}{Dymanus, A.}
\newblock \bibinfo{title}{Hyperfine structure of {CH}$_3${OH}}.
\newblock \emph{\bibinfo{journal}{J. Mol. Spectrosc.}}
  \textbf{\bibinfo{volume}{45}}, \bibinfo{pages}{282 -- 292}
  (\bibinfo{year}{1973}).

\bibitem{belov:16}
\bibinfo{author}{Belov, S.~P.} \emph{et~al.}
\newblock \bibinfo{title}{Torsionally mediated spin-rotation hyperfine
  splittings at moderate to high {J} values in methanol}.
\newblock \emph{\bibinfo{journal}{J. Chem. Phys.}}
  \textbf{\bibinfo{volume}{145}} (\bibinfo{year}{2016}).

\bibitem{coudert:15}
\bibinfo{author}{Coudert, L.}, \bibinfo{author}{Gutl{\'e}, C.},
  \bibinfo{author}{Huet, T.}, \bibinfo{author}{Grabow, J.-U.} \&
  \bibinfo{author}{Levshakov, S.}
\newblock \bibinfo{title}{Spin-torsion effects in the hyperfine structure of
  methanol}.
\newblock \emph{\bibinfo{journal}{J. Chem. Phys.}}
  \textbf{\bibinfo{volume}{143}}, \bibinfo{pages}{044304}
  (\bibinfo{year}{2015}).

\bibitem{eshbach:52}
\bibinfo{author}{Eshbach, J.~R.} \& \bibinfo{author}{Strandberg, M. W.~P.}
\newblock \bibinfo{title}{Rotational magnetic moments of closed shell
  molecules}.
\newblock \emph{\bibinfo{journal}{Phys. Rev.}} \textbf{\bibinfo{volume}{85}},
  \bibinfo{pages}{24--34} (\bibinfo{year}{1952}).

\bibitem{flygare:71}
\bibinfo{author}{Flygare, W.} \& \bibinfo{author}{Benson, R.}
\newblock \bibinfo{title}{The molecular zeeman effect in diamagnetic molecules
  and the determination of molecular magnetic moments ( g values), magnetic
  susceptibilities, and molecular quadrupole moments}.
\newblock \emph{\bibinfo{journal}{Mol. Phys.}} \textbf{\bibinfo{volume}{20}},
  \bibinfo{pages}{225--250} (\bibinfo{year}{1971}).

\bibitem{sutter:76}
\bibinfo{author}{Sutter, D.} \& \bibinfo{author}{Flygare, W.}
\newblock \bibinfo{title}{The molecular {Z}eeman effect}.
\newblock In \bibinfo{editor}{Craig, D.}, \bibinfo{editor}{Mellor, D.},
  \bibinfo{editor}{Gleiter, R.}, \bibinfo{editor}{Gygax, D., R.~Sutter} \&
  \bibinfo{editor}{Flygare, W.} (eds.) \emph{\bibinfo{booktitle}{Bonding
  Structure. Topics in Current Chemistry}}, vol.~\bibinfo{volume}{63},
  \bibinfo{pages}{89--196} (\bibinfo{publisher}{Springer},
  \bibinfo{address}{Berlin, Heidelberg}, \bibinfo{year}{1976}).

\bibitem{flygare:77}
\bibinfo{author}{Flygare, W.}
\newblock \bibinfo{title}{Magnetic interactions in molecules and an analysis of
  molecular electronic charge distribution from magnetic parameters}.
\newblock \emph{\bibinfo{journal}{Chem. Rev.}} \textbf{\bibinfo{volume}{74}},
  \bibinfo{pages}{653--687} (\bibinfo{year}{1974}).

\bibitem{gauss:96a}
\bibinfo{author}{Gauss, J.}, \bibinfo{author}{Ruud, K.} \&
  \bibinfo{author}{Helgaker, T.}
\newblock \bibinfo{title}{Perturbation-dependent atomic orbitals for the
  calculation of spin-rotation constants and rotational g-tensors}.
\newblock \emph{\bibinfo{journal}{J. Chem. Phys.}}
  \textbf{\bibinfo{volume}{105}}, \bibinfo{pages}{2804} (\bibinfo{year}{1996}).

\bibitem{lutnaes:09}
\bibinfo{author}{Lutnaes, O.~B.} \emph{et~al.}
\newblock \bibinfo{title}{Benchmarking density-functional-theory calculations
  of rotational g tensors and magnetizabilities using accurate coupled-cluster
  calculations}.
\newblock \emph{\bibinfo{journal}{J. Chem. Phys.}}
  \textbf{\bibinfo{volume}{131}}, \bibinfo{pages}{144104}
  (\bibinfo{year}{2009}).

\bibitem{cfour}
\bibinfo{author}{Stanton, J.}, \bibinfo{author}{Gauss, J.},
  \bibinfo{author}{M.E., H.} \& \bibinfo{author}{Szalay, P.}
\newblock \bibinfo{title}{{CFOUR}, {C}oupled-{C}luster techniques for
  {C}omputational {C}hemistry}.
\newblock \bibinfo{note}{Http://www.cfour.de}.

\bibitem{dunning:89}
\bibinfo{author}{Dunning~Jr, T.~H.}
\newblock \bibinfo{title}{Gaussian basis sets for use in correlated molecular
  calculations. {I.} {T}he atoms boron through neon and hydrogen}.
\newblock \emph{\bibinfo{journal}{J. Chem. Phys.}}
  \textbf{\bibinfo{volume}{90}}, \bibinfo{pages}{1007--1023}
  (\bibinfo{year}{1989}).

\bibitem{engelbrecht:73}
\bibinfo{author}{Engelbrecht, L.}, \bibinfo{author}{Sutter, D.} \&
  \bibinfo{author}{Dreizier, H.}
\newblock \bibinfo{title}{{Z}eeman effect of molecules with low methyl
  barriers. {I.} {N}itromethane}.
\newblock \emph{\bibinfo{journal}{Z. Naturforsch. A}}
  \textbf{\bibinfo{volume}{28}}, \bibinfo{pages}{709--713}
  (\bibinfo{year}{1973}).

\bibitem{mulliken:55}
\bibinfo{author}{Mulliken, R.~S.}
\newblock \bibinfo{title}{Electronic population analysis on {LCAO}-{MO}
  molecular wave functions. {I}}.
\newblock \emph{\bibinfo{journal}{J. Chem. Phys.}}
  \textbf{\bibinfo{volume}{23}}, \bibinfo{pages}{1833--1840}
  (\bibinfo{year}{1955}).

\bibitem{alexander:85}
\bibinfo{author}{Alexander, M.~H.} \& \bibinfo{author}{Dagdigian, P.~J.}
\newblock \bibinfo{title}{Collision-induced transitions between molecular
  hyperfine levels: Quantum formalism, propensity rules, and experimental study
  of {CaBr} ({X} $^2\sigma_+$) + {Ar}}.
\newblock \emph{\bibinfo{journal}{J. Chem. Phys.}}
  \textbf{\bibinfo{volume}{83}}, \bibinfo{pages}{2191--2200}
  (\bibinfo{year}{1985}).

\bibitem{neufeld:94}
\bibinfo{author}{Neufeld, D.~A.} \& \bibinfo{author}{Green, S.}
\newblock \bibinfo{title}{Excitation of interstellar hydrogen chloride}.
\newblock \emph{\bibinfo{journal}{Astrophys. J.}}
  \textbf{\bibinfo{volume}{432}}, \bibinfo{pages}{158--166}
  (\bibinfo{year}{1994}).

\bibitem{davis:91}
\bibinfo{author}{Davis, S.~L.}
\newblock \bibinfo{title}{Torsionally inelastic collisions between a
  near-symmetric top molecule and a structureless atom}.
\newblock \emph{\bibinfo{journal}{J. Chem. Phys.}}
  \textbf{\bibinfo{volume}{95}}, \bibinfo{pages}{7219--7225}
  (\bibinfo{year}{1991}).

\end{thebibliography}

\clearpage

\begin{figure}[t]
 \centering
 \includegraphics[width=0.45\textwidth]{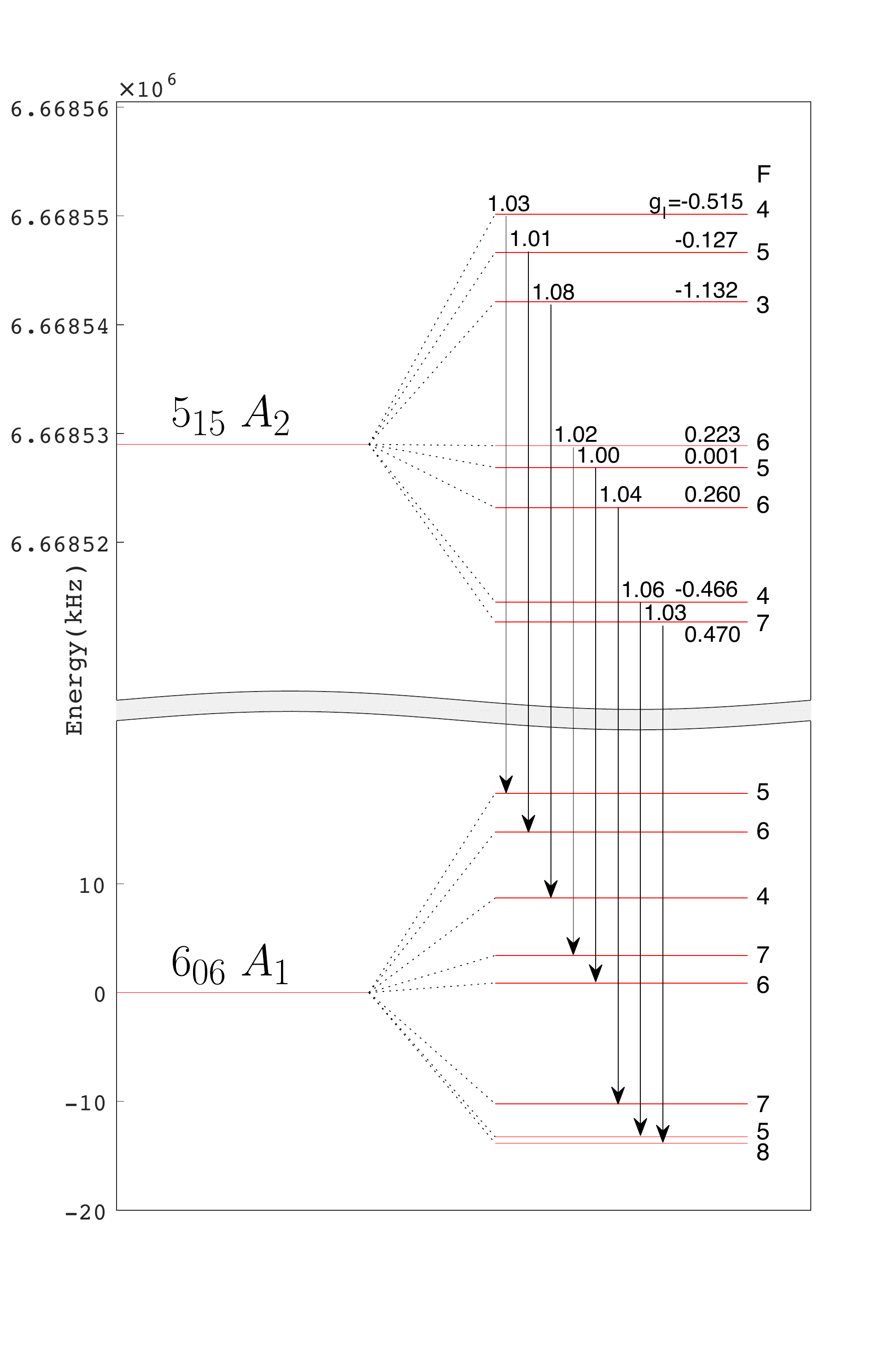}
\caption{Hyperfine structure of the torsion-rotation levels in the 6.7
GHz ($5_{15}\ A_2 \to 6_{06} \ A_1$) transition. The energy of the
$6_{06} \ A_1$ torsion-rotation level is set to zero. Because hyperfine
interactions ($\approx$10 kHz) are much smaller than the
torsion-rotation energy difference ($\approx$10 GHz), we have broken the
y-axis. Torsion-rotation states of $A$-symmetry have nuclear spin
quantum numbers $I$ = 1 and 2, so that for rotational states with $J\geq
2$ there are six levels with $F=J,\ J\pm1$ and two levels with $F=J\pm
2$, each $2F+1$-fold degenerate. Hence, the $F=J,\ J\pm1$ states contain
both $I=1$ and $I=2$ components which are mixed\cite{lankhaar:16}. The
hyperfine structure of the torsion-rotation levels is $\approx 30$ kHz
wide. Arrows indicate the strongest hyperfine transitions with $\Delta F
= \Delta J = 1$, with the Einstein A-coefficients (in $10^{-9}$
s$^{-1}$) indicated above. Land\'{e} g-factors of the transitions in a
magnetic field of 10 mG are given at the righthand side of the upper
energy levels. The rightmost numbers are the $F$ quantum numbers of the
hyperfine states.}
\label{fig:maserline}
\end{figure}

\begin{figure}[t]
  \centering
\includegraphics[width=0.6\textwidth]{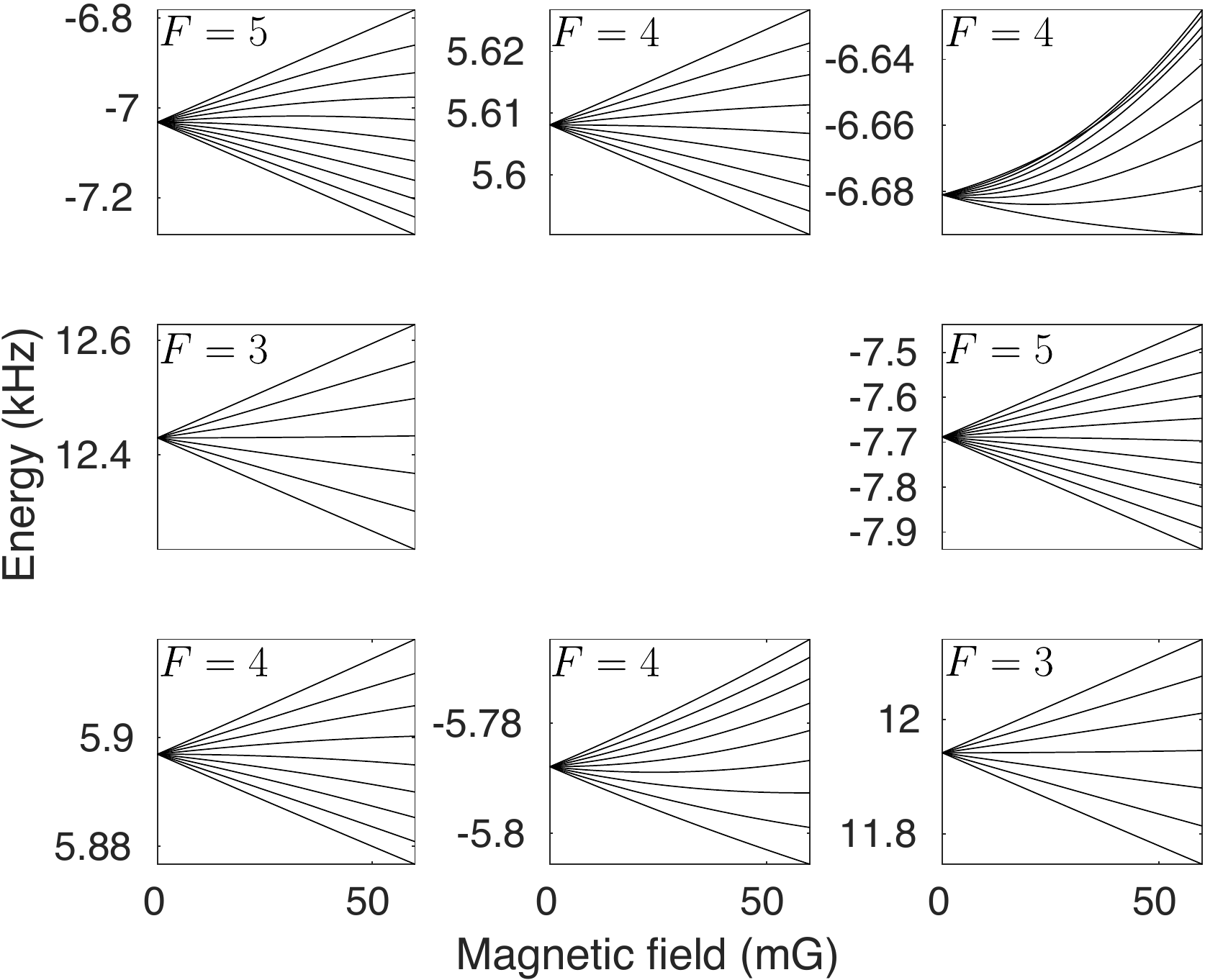}
\caption{Splitting of the 8 hyperfine levels of the torsion-rotation
$4_{-1} \ E$ state as a function of the magnetic field strength. The
quantum number $F$ of each hyperfine level is given in the upper
lefthand corner. In a magnetic field each hyperfine level splits into
$2F+1$ magnetic substates. The energy on the vertical axis is defined
relative to the energy of the corresponding torsion-rotation state
$4_{-1} \ E$.}
\label{fig:j5mag}
\end{figure}

\begin{figure}[h]
 \centering
\includegraphics[width=0.45\textwidth]{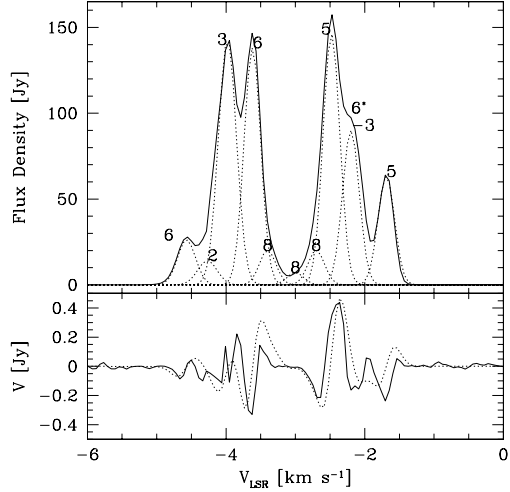}
\caption{Total intensity (flux density in Jy = $10^{-26}$ W m$^{-2}$
Hz$^{-1}$) and circular polarization (V) spectra of the 6.7~GHz
($5_{15}\ A_2 \to 6_{06} \ A_1$) methanol masers around the disc of the
high-mass protostar Cepheus~A~HW2\cite{vlemmings:10}. The spectra were observed with the
Effelsberg 100-m telescope. They originate from methanol
masers with different velocities along the same line of
sight and can be approximated by ten Gaussian peaks. If we assume that a
single hyperfine component is dominant, our calculations imply an
average line-of-sight magnetic field strength $B_{||}=7.7\pm1.0$~mG. The
field strengths (in mG) extracted from the individual components, with
errors of $\approx 20\%$, are indicated in the figure.
This corresponds to a total magnetic
field strength of $|B|=26$~mG.
In one of the
peaks the maser radiation is polarized in the opposite direction than in
the other peaks, which seems to indicate a reverse magnetic field.
We argue, however, that this could be due to pumping of a different
hyperfine component with a different Land\'{e} g-factor. When this
argument holds, the magnetic field (denoted with the asterisk) extracted
from this component has the same direction and is of similar magnitude
as the fields from the other maser components.
Using information on the masing gas and
the mass of the region where the magnetic field is
probed\cite{vlemmings:10}, the recalculated ratio of $\beta=0.2$ between
the thermal and magnetic energy shows the dominance of the magnetic
field. The recalculated mass to magnetic flux ratio compared to the
critical ratio is $\lambda=1.5$.
}
\label{fig:cepheus}
\end{figure}


\end{document}